\documentclass[aip,jmp,preprint]{revtex4-1}

\draft 

\begin{document}

\title{Gauge invariant method for maximum simplification of the field strength in non-Abelian Yang-Mills theories} 

\author{Alcides Garat}
\affiliation{1. Instituto de F\'{\i}sica, Facultad de Ciencias, Igu\'a 4225, esq. Mataojo, Montevideo, Uruguay.}

\date{January 9th, 2012}

\begin{abstract}
A new local gauge invariant method is introduced in order to maximally simplify the expression for a $SU(2)$ non-Abelian field strength. The new tetrads introduced in previous works are going to play a fundamental role in the algorithm presented in this manuscript. Three new local gauge invariant objects are going to guide us through the process of making a field strength block diagonal. The process is also covariant. Any non-trivial isospace field strength projection will become block diagonal through this gauge invariant algorithm. As an application we will find new local observables in Yang-Mills theories.
\end{abstract}



\maketitle 

\section{Introduction}
\label{intro}

In manuscript \cite{A} a covariant method for the local diagonalization of the $U(1)$ electromagnetic stress-energy tensor was presented. At every point in a curved four-dimensional Lorentzian spacetime a new tetrad was introduced for non-null electromagnetic fields such that this tetrad locally and covariantly diagonalizes the stress-energy tensor. At every point the timelike and one spacelike vectors generate a plane that we called blade one \cite{A}$^{,}$\cite{SCH}. The other two spacelike vectors generate a plane that we called blade two. These vectors are constructed with the local extremal field \cite{MW}, its dual, the very metric tensor and a pair of vector fields that represent a generic choice as long as the tetrad vectors do not become trivial. Let us display for the Abelian case the explicit expression for these vectors,

\begin{eqnarray}
U^{\alpha} &=& \xi^{\alpha\lambda}\:\xi_{\rho\lambda}\:X^{\rho} \:
/ \: (\: \sqrt{-Q/2} \: \sqrt{X_{\mu} \ \xi^{\mu\sigma} \
\xi_{\nu\sigma} \ X^{\nu}}\:) \label{U}\\
V^{\alpha} &=& \xi^{\alpha\lambda}\:X_{\lambda} \:
/ \: (\:\sqrt{X_{\mu} \ \xi^{\mu\sigma} \
\xi_{\nu\sigma} \ X^{\nu}}\:) \label{V}\\
Z^{\alpha} &=& \ast \xi^{\alpha\lambda} \: Y_{\lambda} \:
/ \: (\:\sqrt{Y_{\mu}  \ast \xi^{\mu\sigma}
\ast \xi_{\nu\sigma}  Y^{\nu}}\:)
\label{Z}\\
W^{\alpha} &=& \ast \xi^{\alpha\lambda}\: \ast \xi_{\rho\lambda}
\:Y^{\rho} \: / \: (\:\sqrt{-Q/2} \: \sqrt{Y_{\mu}
\ast \xi^{\mu\sigma} \ast \xi_{\nu\sigma} Y^{\nu}}\:) \ .
\label{W}
\end{eqnarray}

We start by stating that at every point in spacetime there is a duality rotation by an angle $-\alpha$ that transforms a non-null electromagnetic field into an extremal field,

\begin{equation}
\xi_{\mu\nu} = e^{-\ast \alpha} f_{\mu\nu}\ = \cos(\alpha)\:f_{\mu\nu} - \sin(\alpha)\:\ast f_{\mu\nu}.\label{dref}
\end{equation}

where  $\ast f_{\mu\nu}={1 \over 2}\:\epsilon_{\mu\nu\sigma\tau}\:f^{\sigma\tau}$ is the dual tensor of $f_{\mu\nu}$. The local scalar $\alpha$ is known as the complexion of the electromagnetic field. It is a local gauge invariant quantity. Extremal fields are essentially electric fields and they satisfy,

\begin{equation}
\xi_{\mu\nu} \ast \xi^{\mu\nu}= 0\ . \label{i0}
\end{equation}

Equation (\ref{i0}) is a condition imposed on (\ref{dref}) and then the explicit expression for the complexion emerges, $\tan(2\alpha) = - f_{\mu\nu}\:\ast f^{\mu\nu} / f_{\lambda\rho}\:f^{\lambda\rho}$. As antisymmetric fields in a four-dimensional Lorentzian spacetime, the extremal fields also verify the identity,

\begin{eqnarray}
\xi_{\mu\alpha}\:\xi^{\nu\alpha} -
\ast \xi_{\mu\alpha}\: \ast \xi^{\nu\alpha} &=& \frac{1}{2}
\: \delta_{\mu}^{\:\:\:\nu}\ Q \ ,\label{i1}
\end{eqnarray}

where $Q=\xi_{\mu\nu}\:\xi^{\mu\nu}=-\sqrt{T_{\mu\nu}T^{\mu\nu}}$
according to equations (39) in \cite{MW}. $Q$ is assumed not to be zero,
because we are dealing with non-null electromagnetic fields. It can be proved that condition (\ref{i0}) and through the use of the general identity,

\begin{eqnarray}
A_{\mu\alpha}\:B^{\nu\alpha} -
\ast B_{\mu\alpha}\: \ast A^{\nu\alpha} &=& \frac{1}{2}
\: \delta_{\mu}^{\:\:\:\nu}\: A_{\alpha\beta}\:B^{\alpha\beta}  \ ,\label{ig}
\end{eqnarray}

which is valid for every pair of antisymmetric tensors in a four-dimensional Lorentzian spacetime \cite{MW}, when applied to the case $A_{\mu\alpha} = \xi_{\mu\alpha}$ and $B^{\nu\alpha} = \ast \xi^{\nu\alpha}$ yields the equivalent condition,

\begin{eqnarray}
\xi_{\alpha\mu}\:\ast \xi^{\mu\nu} &=& 0\ ,\label{i2}
\end{eqnarray}

which is equation (64) in \cite{MW}. The duality rotation given by equation (59) in\cite{MW},

\begin{equation}
f_{\mu\nu} = \xi_{\mu\nu} \: \cos\alpha +
\ast\xi_{\mu\nu} \: \sin\alpha\ ,\label{dr}
\end{equation}

allows us to express the stress-energy tensor in terms of the extremal field,

\begin{equation}
T_{\mu\nu}=\xi_{\mu\lambda}\:\:\xi_{\nu}^{\:\:\:\lambda}
+ \ast \xi_{\mu\lambda}\:\ast \xi_{\nu}^{\:\:\:\lambda}\ .\label{TEMDR}
\end{equation}

With all these elements it becomes trivial to prove that the tetrad \cite{WE}$^{,}$\cite{MTW} (\ref{U}-\ref{W}) is orthonormal and diagonalizes the stress-energy tensor (\ref{TEMDR}). We notice then that we still have to define the vectors $X^{\mu}$ and $Y^{\mu}$. Let us introduce some names. The tetrad vectors have two essential components. For instance in vector $U^{\alpha}$ there are two main structures. First, the skeleton, in this case $\xi^{\alpha\lambda}\:\xi_{\rho\lambda}$, and second, the gauge vector $X^{\rho}$. These do not include the normalization factor $1 / \: (\: \sqrt{-Q/2} \: \sqrt{X_{\mu} \ \xi^{\mu\sigma} \ \xi_{\nu\sigma} \ X^{\nu}}\:)$. The gauge vectors it was proved in manuscript \cite{A} could be anything that does not make the tetrad vectors trivial. That is, the tetrad (\ref{U}-\ref{W}) diagonalizes the stress-energy tensor for any non-trivial gauge vectors $X^{\mu}$ and $Y^{\mu}$. It was therefore proved that we can make different choices for $X^{\mu}$ and $Y^{\mu}$. In geometrodynamics, the Maxwell equations,

\begin{eqnarray}
f^{\mu\nu}_{\:\:\:\:\:;\nu} &=& 0 \label{L1}\nonumber\\
\ast f^{\mu\nu}_{\:\:\:\:\:;\nu} &=& 0 \ , \label{L2}
\end{eqnarray}

are telling us that two potential vector fields $A_{\nu}$ and $\ast A_{\nu}$ exist \cite{CF},

\begin{eqnarray}
f_{\mu\nu} &=& A_{\nu ;\mu} - A_{\mu ;\nu}\label{ER}\nonumber\\
\ast f_{\mu\nu} &=& \ast A_{\nu ;\mu} - \ast A_{\mu ;\nu} \ .\label{DER}
\end{eqnarray}

The $\ast$ in $\ast A_{\nu}$ is just a name, not the Hodge map. The symbol $``;''$ stands for covariant derivative with respect to the metric tensor $g_{\mu\nu}$. We can define then, a tetrad,

\begin{eqnarray}
U^{\alpha} &=& \xi^{\alpha\lambda}\:\xi_{\rho\lambda}\:A^{\rho} \:
/ \: (\: \sqrt{-Q/2} \: \sqrt{A_{\mu} \ \xi^{\mu\sigma} \
\xi_{\nu\sigma} \ A^{\nu}}\:) \label{UO}\\
V^{\alpha} &=& \xi^{\alpha\lambda}\:A_{\lambda} \:
/ \: (\:\sqrt{A_{\mu} \ \xi^{\mu\sigma} \
\xi_{\nu\sigma} \ A^{\nu}}\:) \label{VO}\\
Z^{\alpha} &=& \ast \xi^{\alpha\lambda} \: \ast A_{\lambda} \:
/ \: (\:\sqrt{\ast A_{\mu}  \ast \xi^{\mu\sigma}
\ast \xi_{\nu\sigma}  \ast A^{\nu}}\:)
\label{ZO}\\
W^{\alpha} &=& \ast \xi^{\alpha\lambda}\: \ast \xi_{\rho\lambda}
\:\ast A^{\rho} \: / \: (\:\sqrt{-Q/2} \: \sqrt{\ast A_{\mu}
\ast \xi^{\mu\sigma} \ast \xi_{\nu\sigma} \ast A^{\nu}}\:) \ .
\label{WO}
\end{eqnarray}

The four vectors (\ref{UO}-\ref{WO}) have the following algebraic properties,

\begin{equation}
-U^{\alpha}\:U_{\alpha}=V^{\alpha}\:V_{\alpha}
=Z^{\alpha}\:Z_{\alpha}=W^{\alpha}\:W_{\alpha}=1 \ .\label{TSPAUX}
\end{equation}

Using the equations (\ref{i1}-\ref{i2}) it is simple to prove that (\ref{UO}-\ref{WO}) are orthogonal. When we make the transformation,

\begin{eqnarray}
A_{\alpha} \rightarrow A_{\alpha} + \Lambda_{,\alpha}\ , \label{G1}
\end{eqnarray}

$f_{\mu\nu}$ remains invariant, and the transformation,

\begin{eqnarray}
\ast A_{\alpha} \rightarrow \ast A_{\alpha} +
\ast \Lambda_{,\alpha}\ , \label{G2}
\end{eqnarray}

leaves $\ast f_{\mu\nu}$ invariant,
as long as the functions $\Lambda$ and $\ast \Lambda$ are
scalars. Schouten defined what he called, a two-bladed structure
in a spacetime \cite{SCH}. These blades are the planes determined by the pairs
($U^{\alpha}, V^{\alpha}$) and ($Z^{\alpha}, W^{\alpha}$).
It was proved in \cite{A} that the transformation (\ref{G1}) generates a ``rotation'' of the tetrad vectors ($U^{\alpha}, V^{\alpha}$) into ($\tilde{U}^{\alpha}, \tilde{V}^{\alpha}$) such that these ``rotated'' vectors ($\tilde{U}^{\alpha}, \tilde{V}^{\alpha}$) remain in the plane or blade one generated by ($U^{\alpha}, V^{\alpha}$). It was also proved in \cite{A} that the transformation (\ref{G2}) generates a ``rotation'' of the tetrad vectors ($Z^{\alpha}, W^{\alpha}$) into ($\tilde{Z}^{\alpha}, \tilde{W}^{\alpha}$) such that these ``rotated'' vectors ($\tilde{Z}^{\alpha}, \tilde{W}^{\alpha}$) remain in the plane or blade two generated by ($Z^{\alpha}, W^{\alpha}$).  For the sake of simplicity we are going to assume that the transformation of the two vectors $(U^{\alpha},\:V^{\alpha})$ on blade one, given in (\ref{UO}-\ref{VO}), by the ``angle'' $\phi$ is a proper transformation, that is, a boost. For discrete improper transformations the result follows the same lines \cite{A}. Therefore we can write,

\begin{eqnarray}
U^{\alpha}_{(\phi)}  &=& \cosh(\phi)\: U^{\alpha} +  \sinh(\phi)\: V^{\alpha} \label{UT} \\
V^{\alpha}_{(\phi)} &=& \sinh(\phi)\: U^{\alpha} +  \cosh(\phi)\: V^{\alpha} \label{VT} \ .
\end{eqnarray}

The transformation of the two tetrad vectors $(Z^{\alpha},\:W^{\alpha})$ on blade two, given in (\ref{ZO}-\ref{WO}), by the ``angle'' $\varphi$, can be expressed as,

\begin{eqnarray}
Z^{\alpha}_{(\varphi)}  &=& \cos(\varphi)\: Z^{\alpha} -  \sin(\varphi)\: W^{\alpha} \label{ZT} \\
W^{\alpha}_{(\varphi)}  &=& \sin(\varphi)\: Z^{\alpha} +  \cos(\varphi)\: W^{\alpha} \label{WT} \ .
\end{eqnarray}

It is a simple exercise in algebra to see that the equalities $U^{[\alpha}_{(\phi)}\:V^{\beta]}_{(\phi)} = U^{[\alpha}\:V^{\beta]}$ and $Z^{[\alpha}_{(\varphi)}\:W^{\beta]}_{(\varphi)} = Z^{[\alpha}\:W^{\beta]}$ are true. These equalities are telling us that these antisymmetric tetrad objects are gauge invariant. We remind ourselves that it was proved in manuscript \cite{A} that the group of local electromagnetic gauge transformations is isomorphic to the group LB1 of boosts plus discrete transformations on blade one, and independently to LB2, the group of spatial rotations on blade two. Equations (\ref{UT}-\ref{VT}) represent a local electromagnetic gauge transformation of the vectors $(U^{\alpha}, V^{\alpha})$. Equations (\ref{ZT}-\ref{WT}) represent a local electromagnetic gauge transformation of the vectors $(Z^{\alpha}, W^{\alpha})$. Written in terms of these tetrad vectors, the electromagnetic field is,

\begin{equation}
f_{\alpha\beta} = -2\:\sqrt{-Q/2}\:\:\cos\alpha\:\:U_{[\alpha}\:V_{\beta]} +
2\:\sqrt{-Q/2}\:\:\sin\alpha\:\:Z_{[\alpha}\:W_{\beta]}\ .\label{EMF}
\end{equation}

Equation (\ref{EMF}) represents maximum simplification in the expression of the electromagnetic field. The true degrees of freedom are the local scalars $\sqrt{-Q/2}$ and $\alpha$. Local gauge invariance is manifested explicitly through the possibility of ``rotating'' through a scalar angle $\phi$ on blade one by a local gauge transformation (\ref{UT}-\ref{VT}) the tetrad vectors $U^{\alpha}$ and $V^{\alpha}$, such that
$U_{[\alpha}\:V_{\beta]}$ remains invariant \cite{A}. Analogous for discrete transformations on blade one. Similar analysis on blade two. A spatial ``rotation'' of the tetrad vectors $Z^{\alpha}$ and $W^{\alpha}$ through an ``angle'' $\varphi$ as in (\ref{ZT}-\ref{WT}), such that $Z_{[\alpha}\:W_{\beta]}$ remains invariant \cite{A}. All this formalism clearly provides a technique to maximally simplify the expression for the electromagnetic field strength. It is block diagonalized automatically by the tetrad (\ref{UO}-\ref{WO}). This is not the case for the non-Abelian $SU(2)$ field strength. We do not have an automatic block diagonalization. We have to develop a new algorithm. This is the goal of this note. In section \ref{nonabeltetrads} we are going to introduce appropriate tetrads for the $SU(2)$ non-Abelian Yang-Mills case. In section \ref{Gauge invariants} we are going to introduce three new gauge invariant objects for the non-Abelian local $SU(2)$ case. In section \ref{diagonal} we are going to introduce the new algorithm for local block diagonalization of the non-Abelian field strength. Finally, in section \ref{application:observables} we are going to introduce as an application, new observables found with our new tetrads and our new method to locally block diagonalize the field strength in non-Abelian Yang-Mills theories. Throughout the paper we use the conventions of manuscript \cite{MW}. In particular we use a metric with sign conventions -+++. The only difference in notation with \cite{MW} will be that we will call our geometrized electromagnetic potential $A^{\alpha}$, where $f_{\mu\nu}=A_{\nu ;\mu} - A_{\mu ;\nu}$ is the geometrized electromagnetic field $f_{\mu\nu}= (G^{1/2} / c^2) \: F_{\mu\nu}$. Analogously, $f^{k}_{\mu\nu}$ are the geometrized Yang-Mills field components, $f^{k}_{\mu\nu}= (G^{1/2} / c^2) \: F^{k}_{\mu\nu}$.

\section{Tetrads for non-Abelian theories}
\label{nonabeltetrads}

This section has an illustrative purpose, fundamentally. We will show many ways to construct local $SU(2)$ gauge invariant skeletons. Later, in section \ref{diagonal} we will use a particular and convenient way of constructing a skeleton adapted to the purpose of block diagonalizing the field strength isospace projections. Let us define then, an extremal field for non-Abelian theories as,
\begin{equation}
\zeta_{\mu\nu} = \cos\beta \:\: f_{\mu\nu}-
\sin\beta \:\: \ast f_{\mu\nu} \ ,\label{exsu2}
\end{equation}

In order to define the complexion $\beta$, we are going to impose the $SU(2)$ invariant condition,

\begin{eqnarray}
Tr[\zeta_{\mu\nu}\:\ast \zeta^{\mu\nu}]=\zeta^{k}_{\mu\nu}\:\ast \zeta^{k\mu\nu} &=& 0\ ,\label{ccsu2}
\end{eqnarray}

where the summation convention was applied on the internal index $k$. The complexion condition (\ref{ccsu2}) is not an additional condition for the field strength. We are just using a generalized duality transformation, and defining through it this new local scalar complexion $\beta$. After the fields are available from the equations, not before. We simply generalized the definition for the Abelian complexion, found through a duality transformation as well. Then, the local $SU(2)$ invariant complexion $\beta$ turns out to be,

\begin{eqnarray}
\tan(2\beta) = - f^{k}_{\mu\nu}\:\ast f^{k\mu\nu} / f^{p}_{\lambda\rho}\:f^{p\lambda\rho}\ ,\label{compksu2}
\end{eqnarray}

where again the summation convention was applied on both $k$ and $p$.

Now we would like to consider gauge covariant derivatives. For instance, the gauge covariant derivatives of the three extremal field internal components,

\begin{eqnarray}
\zeta_{k\mu\nu\mid\rho} = \zeta_{k\mu\nu\, ; \, \rho} + g \: \epsilon_{klp}\: A_{l\rho}\:\zeta_{p\mu\nu}\ .\label{gcd}
\end{eqnarray}

where $\epsilon_{klp}$ is the completely skew-symmetric tensor in three dimensions with $\epsilon_{123} = 1$, and $g$ is the coupling constant. The symbol ``;'' stands for the usual covariant derivative associated with the metric tensor $g_{\mu\nu}$. If we consider for instance the Einstein-Maxwell-Yang-Mills vacuum field equations,

\begin{eqnarray}
R_{\mu\nu} &=& T^{(ym)}_{\mu\nu} + T^{(em)}_{\mu\nu}\label{eyme}\\
f^{\mu\nu}_{\:\:\:\:\:;\nu} &=& 0 \label{EM1}\\
\ast f^{\mu\nu}_{\:\:\:\:\:;\nu} &=& 0 \label{EM2}\\
f^{k\mu\nu}_{\:\:\:\:\:\:\:\:\mid \nu} &=& 0 \label{ymvfe1}\\
\ast f^{k\mu\nu}_{\:\:\:\:\:\:\:\:\mid \nu} &=& 0 \ . \label{ymvfe2}
\end{eqnarray}

The field equations (\ref{EM1}-\ref{EM2}) provide a hint about the existence of two electromagnetic field potentials, as said in the first paper \cite{A} ``Tetrads in geometrodynamics'', not independent from each other, but due to the symmetry of the equations, available for our construction. $A^{\mu}$ and $\ast A^{\mu}$ are the two electromagnetic potentials. $\ast A^{\mu}$ is therefore a name, we are not using the Hodge map at all in this case. These two potentials are not independent from each other, nonetheless they exist and are available for our construction. Similar for the two Non-Abelian  equations (\ref{ymvfe1}-\ref{ymvfe2}). The Non-Abelian potential $A^{k\mu}$ is available for our construction as well \cite{MC}$^{,}$\cite{YM}$^{,}$\cite{RU}.
With all these elements, we can proceed as an example, to define the antisymmetric field,

\begin{eqnarray}
\omega_{\mu\nu} = (\zeta^{p}_{\:\sigma\tau}\: \zeta^{p}_{\mu\nu})\:(\zeta^{k\sigma\rho}_{\:\:\:\:\:\:\:\:\mid\rho}\:\ast \zeta^{k\tau\lambda}_{\:\:\:\:\:\:\:\:\mid\lambda}) \ .\label{anti1}
\end{eqnarray}

This particular intermediate field in our construction could also be chosen to be,

\begin{eqnarray}
\omega_{\mu\nu} = (\ast \zeta^{p}_{\:\sigma\tau}\: \ast\zeta^{p}_{\mu\nu})\:\left(\zeta^{k\sigma\rho}\:\ast \zeta^{k\tau\lambda}-\ast\zeta^{k\sigma\rho}\:\zeta^{k\tau\lambda}\right)\:T_{\rho\lambda} \ .\label{anti2}
\end{eqnarray}

There are many possible choices for this intermediate field $\omega_{\mu\nu}$, we are just showing two of them. The summation convention on the internal index $k$ as well as $p$ was applied. It is clear that (\ref{anti1}) or (\ref{anti2}) are invariant under $SU(2)$ local gauge transformations. Expressions (\ref{anti1}) or (\ref{anti2}) are nothing but explicit examples among many. Once our choice is made, then the duality rotation we perform next, in order to obtain the new extremal field is,


\begin{eqnarray}
\epsilon_{\mu\nu} = \cos\vartheta \: \omega_{\mu\nu} - \sin\vartheta \:
\ast \omega_{\mu\nu}\ .\label{extremalR}
\end{eqnarray}

As always we choose this complexion $\vartheta$ to be defined by the condition,

\begin{eqnarray}
\epsilon_{\mu\nu}\:\ast \epsilon^{\mu\nu} &=& 0\ ,\label{rc}
\end{eqnarray}

which implies that,

\begin{eqnarray}
\tan(2\vartheta) = - \omega_{\mu\nu}\:\ast \omega^{\mu\nu} / \omega_{\lambda\rho}\:\omega^{\lambda\rho}\ .\label{compr}
\end{eqnarray}

This new kind of local $SU(2)$ gauge invariant extremal tensor $\epsilon_{\mu\nu}$, allows in turn for the construction of the new tetrad,

\begin{eqnarray}
S_{(1)}^{\mu} &=& \epsilon^{\mu\lambda}\:\epsilon_{\rho\lambda}\:X^{\rho}
\label{S1}\\
S_{(2)}^{\mu} &=& \sqrt{-Q_{ym}/2} \: \epsilon^{\mu\lambda} \: X_{\lambda}
\label{S2}\\
S_{(3)}^{\mu} &=& \sqrt{-Q_{ym}/2} \: \ast \epsilon^{\mu\lambda} \: Y_{\lambda}
\label{S3}\\
S_{(4)}^{\mu} &=& \ast \epsilon^{\mu\lambda}\: \ast\epsilon_{\rho\lambda}
\:Y^{\rho}\ ,\label{S4}
\end{eqnarray}

where $Q_{ym} = \epsilon_{\mu\nu}\:\epsilon^{\mu\nu}$. With the help of identity (\ref{ig}), when applied to the case $A_{\mu\alpha} = \epsilon_{\mu\alpha}$ and $B^{\nu\alpha} = \ast \epsilon^{\nu\alpha}$ yields the equivalent condition,

\begin{eqnarray}
\epsilon_{\alpha\nu}\:\ast \epsilon^{\mu\nu} &=& 0\ ,\label{isu2}
\end{eqnarray}

It is straightforward using (\ref{ig}) for $A_{\mu\alpha} = \epsilon_{\mu\alpha}$ and $B^{\nu\alpha} = \epsilon^{\nu\alpha}$, and (\ref{isu2}), to prove that vectors (\ref{S1}-\ref{S4}) are orthogonal. We are going to call for future reference for instance $\epsilon^{\mu\lambda}\:\epsilon_{\rho\lambda}$ the skeleton of the tetrad vector $S_{(1)}^{\mu}$, and $X^{\rho}$ the gauge vector. In the case of $S_{(3)}^{\mu}$, the skeleton will be $\ast \epsilon^{\mu\lambda}$, and $Y_{\lambda}$ will be the gauge vector. It is clear now that skeletons are gauge invariant. This property guarantees that the vectors under local $U(1)$ or $SU(2)$ gauge transformations are not going to leave their original planes or blades, keeping therefore the metric tensor explicitly invariant.

The question remains about the choice that we can make for the two gauge vector fields
$X^{\sigma}$ and $Y^{\sigma}$ in (\ref{S1}-\ref{S4}) such that we can reproduce in the $SU(2)$ environment, the tetrad transformation properties of the Abelian environment. One possible choice could be $X^{\sigma} = Y^{\sigma} = Tr[\Sigma^{\alpha\beta}\:E_{\alpha}^{\:\:\rho}\: E_{\beta}^{\:\:\lambda}\:\ast \xi_{\rho}^{\:\:\sigma}\:\ast \xi_{\lambda\tau}\:A^{\tau}]$. The nature of the object $\Sigma^{\alpha\beta}$ is explained in section \ref{sec:appI}. $E_{\alpha}^{\:\:\rho}$ are tetrad vectors that transform from a locally inertial coordinate system, into a general curvilinear coordinate system. From now on, Greek indices $\alpha$, $\beta$, $\delta$, $\epsilon$, $\gamma$, and $\kappa$, will be reserved for locally inertial coordinate systems. There is a particular explicit choice that we can make for these tetrads $E_{\alpha}^{\:\:\rho}$. We can choose the tetrad vectors we already know from \cite{A}, for electromagnetic fields in curved space-times. Following the same notation in \cite{A}, we call $E_{o}^{\:\:\rho} = U^{\rho}$, $E_{1}^{\:\:\rho} = V^{\rho}$, $E_{2}^{\:\:\rho} = Z^{\rho}$, $E_{3}^{\:\:\rho} = W^{\rho}$. The electromagnetic extremal tensor $\xi_{\rho\sigma}$, and its dual $\ast \xi_{\rho\sigma}$ are also already known from \cite{A}. That is, we are making use of the already defined tetrads built for space-times where electromagnetic fields are present, in order to allow for the use of the object $\Sigma^{\alpha\beta}$ which is key in our construction. The key lies in the translating quality of this object between $SU(2)$ local gauge transformations and local Lorentz transformations. We would like to consider one more property of these chosen vector fields $X^{\rho}$ and $Y^{\rho}$. The structure $E_{\alpha}^{\:\:[\rho}\:E_{\beta}^{\:\:\lambda]}\:\ast \xi_{\rho\sigma}\:\ast \xi_{\lambda\tau}$ is invariant under $U(1)$ local gauge transformations. Essentially, because of the electromagnetic extremal field property \cite{A}$^{,}$\cite{MW}, $\xi_{\mu\sigma}\:\ast \xi^{\mu\tau} = 0$. In the covariant expression $E_{\alpha}^{\:\:[\rho}\:E_{\beta}^{\:\:\lambda]}\:\ast \xi_{\rho\sigma}\:\ast \xi_{\lambda\tau}$ only the vectors with $\alpha = 2, 3$ and $\beta = 2, 3$ survive, see equations (\ref{U}-\ref{W}) and equation (\ref{i2}) plus the $U(1)$ local gauge transformation tetrad invariant $Z^{[\rho}_{(\varphi)}\:W^{\lambda]}_{(\varphi)} = Z^{[\rho}\:W^{\lambda]}$.

\section{Gauge invariants}
\label{Gauge invariants}

First of all we would like to introduce new gauge invariant objects built out of the tetrad components of the field strength tensor. Given the tetrad $W_{(o)}^{\mu}$, $W_{(1)}^{\mu}$, $W_{(2)}^{\mu}$, $W_{(3)}^{\mu}$, (no confusion should arise with vector $E_{3}^{\:\:\rho} = W^{\rho}$ which is just one vector in the electromagnetic tetrad) which we consider to be the normalized version of $S_{(1)}^{\mu}$, $S_{(2)}^{\mu}$, $S_{(3)}^{\mu}$, $S_{(4)}^{\mu}$, we perform the gauge transformations on blades one and two,

\begin{eqnarray}
\tilde{W}_{(o)}^{\mu} &=& \cosh\phi\:W_{(o)}^{\mu} + \sinh\phi\:W_{(1)}^{\mu}\label{GT1}\\
\tilde{W}_{(1)}^{\mu} &=& \sinh\phi\:W_{(o)}^{\mu} + \cosh\phi\:W_{(1)}^{\mu}\label{GT2}\\
\tilde{W}_{(2)}^{\mu} &=& \cos\psi\:W_{(2)}^{\mu} - \sin\psi\:W_{(3)}^{\mu}\label{GT3}\\
\tilde{W}_{(3)}^{\mu} &=& \sin\psi\:W_{(2)}^{\mu} + \cos\psi\:W_{(3)}^{\mu}\ . \label{GT4}
\end{eqnarray}

That equations (\ref{GT1}-\ref{GT2}) are the result of a local $SU(2)$ gauge transformation on blade one at every point was proven in reference \cite{A2}. Similar for equations (\ref{GT3}-\ref{GT4}) on blade two. It was also proven there that the local group of $SU(2)$ gauge transformations is isomorphic to the triple tensor product $(\bigotimes LB1)^{3}$ and independently also to $(\bigotimes LB2)^{3}$ see manuscript \cite{A2}. Then, it is a matter of algebra to prove that the following objects are invariant under the set of transformations (\ref{GT1}-\ref{GT4}),

\begin{eqnarray}
\lefteqn{ \left(\:W_{(0)}^{\mu}\:\overline{f}_{\mu\nu}\:W_{(1)}^{\nu}\right)\:\left[W_{(0)}^{\lambda}\:W_{(1)}^{\rho} - W_{(0)}^{\rho}\:W_{(1)}^{\lambda}\right] } \label{GI1} \\
&&\left(\:W_{(0)}^{\mu}\:\overline{f}_{\mu\nu}\:W_{(2)}^{\nu}\right)\:\left[W_{(0)}^{\lambda}\:W_{(2)}^{\rho}\right]
+\left(\:W_{(0)}^{\mu}\:\overline{f}_{\mu\nu}\:W_{(3)}^{\nu}\right)\:\left[W_{(0)}^{\lambda}\:W_{(3)}^{\rho}\right] - \nonumber \\
&&\left(\:W_{(1)}^{\mu}\:\overline{f}_{\mu\nu}\:W_{(2)}^{\nu}\right)\:\left[W_{(1)}^{\lambda}\:W_{(2)}^{\rho}\right] -
\left(\:W_{(1)}^{\mu}\:\overline{f}_{\mu\nu}\:W_{(3)}^{\nu}\right)\:\left[W_{(1)}^{\lambda}\:W_{(3)}^{\rho}\right] \label{GI2}\\
&&\left(\:W_{(2)}^{\mu}\:\overline{f}_{\mu\nu}\:W_{(3)}^{\nu}\right)\:\left[W_{(2)}^{\lambda}\:W_{(3)}^{\rho} - W_{(2)}^{\rho}\:W_{(3)}^{\lambda}\right] \ ,\label{GI3}
\end{eqnarray}

where $f_{\mu\nu} = f^{a}_{\mu\nu}\:\sigma^{a}$, and $\vec{n} = n^{a}\:\sigma^{a}$  are vectors in isospace. The $\cdot$ means product in isospace. $\sigma^{a}$ are the Pauli matrices (see section \ref{sec:appI}) and the summation convention is applied on the internal index $a$. In equations (\ref{GI1}-\ref{GI3}) by $\overline{f}_{\mu\nu}$ we mean $Tr[\vec{n}\: \cdot \: f_{\mu\nu}] = n^{a}\:f^{a}_{\mu\nu}$ where again the summation convention is applied on the internal index $a$. The vector components are defined as,

\begin{eqnarray}
\vec{n} = (\cos\theta_{1},\cos\theta_{2},\cos\theta_{3})  \label{ISO}
\end{eqnarray}

where all the three isoangles are local scalars that satisfy,

\begin{eqnarray}
\Sigma_{a=1}^{3} \cos^{2}\theta_{a} = 1  \label{ISOSUM}
\end{eqnarray}

In isospace $\vec{n} = n^{a}\:\sigma^{a}$ transforms under a local $SU(2)$ gauge transformation $S$, as $S^{-1}\:\vec{n}\:S$, see chapter III in \cite{CBDW} and also reference \cite{GRSYMM}. The tensor $f_{\mu\nu} = f^{a}_{\mu\nu}\:\sigma^{a}$ is going to transform as $f_{\mu\nu} \rightarrow S^{-1}\:f_{\mu\nu}\:S$. Therefore, $\overline{f}_{\mu\nu}$ which is nothing but compact notation for $Tr[\vec{n}\: \cdot \: f_{\mu\nu}]$ is a local $SU(2)$ gauge invariant object. The subtlety here is the following. Using any normalized tetrads, and under tetrad transformations of the kind (\ref{GT1}-\ref{GT4}), the objects (\ref{GI1}-\ref{GI3}) are going to remain invariant. The point is that the transformations (\ref{GT1}-\ref{GT4}), are $SU(2)$ local tetrad gauge transformations, or tetrad gauge generated trasformations, see manuscripts \cite{A}$^{,}$\cite{A2}. It is the way in which the normalized version of tetrad vectors (\ref{S1}-\ref{S4}) transform on blades one and two under locally generated $SU(2)$ gauge transformations. The tensor $\overline{f}_{\mu\nu} = Tr[\vec{n}\: \cdot \:f_{\mu\nu}]$ is gauge invariant by itself as we already know. Then these are true new gauge invariants under (\ref{GT1}-\ref{GT4}). We might wonder what happens with the objects (\ref{GI1}-\ref{GI3}), when we perform discrete gauge transformations on blade one. It is evident that all of the objects remain invariant under a tetrad full inversion on blade one. However, under the discrete transformation represented by equations (64-65) in reference \cite{A}, while objects (\ref{GI1}) and (\ref{GI3}) remain invariant, object (\ref{GI2}) changes in a global sign (gets multiplied globally by $-1$). Therefore we can say that objects (\ref{GI1}) and (\ref{GI3}) are true and new gauge invariants, while object (\ref{GI2}) is invariant under boosts generated gauge transformations on blade one, rotations on blade two, full inversions on blade one, but gets multiplied by $-1$ under the discrete gauge generated transformation on blade one given by equations (64-65) in reference \cite{A}. We are going to make use of these gauge invariant properties of objects (\ref{GI1}-\ref{GI3}) in the next section that deals with the block diagonalization of the field strength tensor.

\section{Block diagonalization of the field strength tensor}
\label{diagonal}

We proceed now to extend to the non-Abelian case the algorithm for the local block diagonalization of the field strength tensor. The Abelian case was studied thoroughly in manuscript \cite{A}. In the previous section \ref{Gauge invariants} we found that we can build with the field strength tensor and the new tetrads, three objects that are locally gauge invariant. This is a mathematical truth that can be easily checked. Then, we might ask about the usefulness of the existence of these three new local gauge invariant objects, and our answer is the following. These three new local gauge invariant objects allow us to connect gauge invariance with three different blocks in the field strength tensor, one block off-diagonal and two diagonal blocks, separately. By field strength tensor we mean as in the previous section \ref{Gauge invariants} the object $Tr[\vec{n}\: \cdot \: f_{\mu\nu}] = n^{a}\:f^{a}_{\mu\nu}$ that we expressed compactly as $\overline{f}_{\mu\nu}$. It must be clear from the outset that the object that we are going to block diagonalize locally is $\overline{f}_{\mu\nu} = Tr[\vec{n}\: \cdot \: f_{\mu\nu}] = n^{a}\:f^{a}_{\mu\nu}$ such that the local isovector $\vec{n}$ will remain unchanged during the process. It is given at the outset of the algorithm. We are block diagonalizing isospace projections of the field strength. Therefore these three new gauge invariant objects are going to guide us in establishing a local gauge invariant process of block diagonalization of the field strength tensor. Their existence means that we can block diagonalize the field strength tensor in a gauge invariant way, locally. We start to develop this new technique by putting forward a generalized duality transformation for non-Abelian fields. For instance we might choose,

\begin{equation}
\varepsilon_{\mu\nu} =  Tr[\vec{m}\: \cdot \: f_{\mu\nu} - \vec{l}\: \cdot \: \ast f_{\mu\nu}] \ ,\label{gendual}
\end{equation}

where $f_{\mu\nu} = f^{a}_{\mu\nu}\:\sigma^{a}$, $\vec{m} = m^{a}\:\sigma^{a}$ and $\vec{l} = l^{a}\:\sigma^{a}$ are vectors in isospace. The $\cdot$ means again product in isospace. Once more we stress that $\sigma^{a}$ are the Pauli matrices (see section \ref{sec:appI}) and the summation convention is applied on the internal index $a$. The vector components are defined as,

\begin{eqnarray}
\lefteqn{ \vec{m} = (\cos\alpha_{1},\cos\alpha_{2},\cos\alpha_{3}) } \label{ISO1} \\
&&\vec{l} = (\cos\beta_{1},\cos\beta_{2},\cos\beta_{3}) \ , \label{ISO2}
\end{eqnarray}

where all the six isoangles are local scalars that satisfy,

\begin{eqnarray}
\lefteqn{ \Sigma_{a=1}^{3} \cos^{2}\alpha_{a} = 1 } \label{ISOSUM1} \\
&&\Sigma_{a=1}^{3} \cos^{2}\beta_{a} = 1 \ . \label{ISOSUM2}
\end{eqnarray}

In isospace $\vec{m} = m^{a}\:\sigma^{a}$ transforms under a local $SU(2)$ gauge transformation $S$, as $S^{-1}\:\vec{m}\:S$, see chapter III in \cite{CBDW} and also reference \cite{GRSYMM}, and similar for $\vec{l} = l^{a}\:\sigma^{a}$. The tensor $f_{\mu\nu} = f^{a}_{\mu\nu}\:\sigma^{a}$ transforms as
$f_{\mu\nu} \rightarrow S^{-1}\:f_{\mu\nu}\:S$. Therefore $\varepsilon_{\mu\nu}$ is manifestly gauge invariant. We can see from (\ref{ISO1}-\ref{ISO2}) and (\ref{ISOSUM1}-\ref{ISOSUM2}) that only four of the six angles in isospace are independent. Next we perform one more duality transformation,

\begin{equation}
\Omega_{\mu\nu} = \cos\alpha_{d} \:\: \varepsilon_{\mu\nu} -
\sin\alpha_{d} \:\: \ast \varepsilon_{\mu\nu} \ ,\label{diagdual}
\end{equation}

such that the complexion $\alpha_{d}$ is defined by the usual local condition $\Omega_{\mu\nu}\:\ast \Omega^{\mu\nu} = 0$, see section \ref{intro} and reference \cite{A},

\begin{eqnarray}
\tan(2\alpha_{d}) = - \varepsilon_{\mu\nu}\:\ast \varepsilon^{\mu\nu} / \varepsilon_{\lambda\rho}\:\varepsilon^{\lambda\rho}\ .\label{compdd}
\end{eqnarray}

All the conclusions derived in \cite{A} are valid in this context and therefore exactly as in reference \cite{A}. Using the local antisymmetric tensor $\Omega_{\mu\nu}$, we can produce tetrad skeletons and with new gauge vectors $X_{d}^{\sigma}$ and $Y_{d}^{\sigma}$ we can build a new normalized tetrad. This new tetrad that we call $T_{\alpha}^{\mu}$ has four independent isoangles included in its definition, in the skeletons. There is also the freedom to introduce an LB1 and an LB2 local  $SU(2)$ generated transformations on both blades by new angles $\phi_{d}$ and $\psi_{d}$ (through the gauge vectors $X_{d}^{\sigma}$ and $Y_{d}^{\sigma}$) which are not yet fixed and represent two more independent angles that are not going to intervene in our algorithm. Essentially because objects (\ref{GI1}) and (\ref{GI3}) are local gauge invariants, and therefore any non-trivial choice for $X_{d}^{\sigma}$ and $Y_{d}^{\sigma}$ is good and makes no difference.  Having four independent and undefined angles, we are going to use this freedom to choose them when fixing the four block diagonalization conditions for the field strength tensor. It must be highlighted and stressed that since the local antisymmetric tensor $\Omega_{\mu\nu}$ is $SU(2)$ gauge invariant, then the tetrad vectors skeletons are $SU(2)$ gauge invariant. This was a fundamental condition that we made in previous sections in order to ensure the metric invariance when performing LB1 and LB2 transformations. Then, we proceed to impose the block diagonalization conditions,

\begin{eqnarray}
\lefteqn{
\overline{f}_{o2} = T_{o}^{\mu}\:\overline{f}_{\mu\nu}\: T_{2}^{\nu} = 0} \label{diagcond2} \\
&&\overline{f}_{o3} = T_{o}^{\mu}\:\overline{f}_{\mu\nu}\:T_{3}^{\nu} = 0 \label{diagcond3} \\
&&\overline{f}_{12} = T_{1}^{\mu}\:\overline{f}_{\mu\nu}\: T_{2}^{\nu} = 0 \label{diagcond4} \\
&&\overline{f}_{13} = T_{1}^{\mu}\:\overline{f}_{\mu\nu}\: T_{3}^{\nu} = 0 \ . \label{diagcond5}
\end{eqnarray}

These are finally the four equations that locally define the four angles $\alpha_{1},\:\alpha_{2},\:\beta_{1},\:\beta_{2}$, for instance. The other two $\alpha_{3},\:\beta_{3}$ are determined by equations (\ref{ISOSUM1}-\ref{ISOSUM2}) once the other four have already been determined through equations (\ref{diagcond2}-\ref{diagcond5}). Once the field strength tensor has been block diagonalized, always assuming that the local diagonalization process is possible, we can study the gauge invariants (\ref{GI1}-\ref{GI3}). We imposed the off-diagonal tetrad components of the field strength tensor (\ref{diagcond2}-\ref{diagcond5}) to be zero. These four equations are manifestly and locally $SU(2)$ gauge invariant by themselves under LB1 and LB2 local transformations of the vectors $T_{\alpha}^{\mu}$, analogous to transformations (\ref{GT1}-\ref{GT4}). As an example let us see for instance the local gauge transformation of the object $T_{(0)}^{\mu}\:\overline{f}_{\mu\nu}\:T_{(2)}^{\nu}$. That is, under local transformations like (\ref{GT1}-\ref{GT4}) the object $T_{(0)}^{\mu}\:\overline{f}_{\mu\nu}\:T_{(2)}^{\nu} = Tr[\vec{n}\: \cdot \: f_{\mu\nu}]\:T_{(0)}^{\mu}\:T_{(2)}^{\nu}$ goes into,

\begin{eqnarray}
&&Tr[\vec{n}\: \cdot \: f_{\mu\nu}]\:[\cosh\phi\:T_{(0)}^{\mu} + \sinh\phi\:T_{(1)}^{\mu}]\:[\cos\varphi\:T_{(2)}^{\mu} - \sin\varphi\:T_{(3)}^{\mu}] = \nonumber \\ &&\cosh\phi\:\cos\varphi\:Tr[\vec{n}\: \cdot \: f_{\mu\nu}]\:T_{(0)}^{\mu}\:T_{(2)}^{\mu} -  \nonumber \\
&&\cosh\phi\:\sin\varphi\:Tr[\vec{n}\: \cdot \: f_{\mu\nu}]\:T_{(0)}^{\mu}\:T_{(3)}^{\mu} + \nonumber \\
&&\sinh\phi\:\cos\varphi\:Tr[\vec{n}\: \cdot \: f_{\mu\nu}]\:T_{(1)}^{\mu}\:T_{(2)}^{\mu} - \nonumber \\
&&\sinh\phi\:\sin\varphi\:Tr[\vec{n}\: \cdot \: f_{\mu\nu}]\:T_{(1)}^{\mu}\:T_{(3)}^{\mu}\ .\label{locgaugetr02}
\end{eqnarray}

If in one local gauge, equations (\ref{diagcond2}-\ref{diagcond5}) are satisfied, then, in any new local gauge they also will be. Therefore, the new off-diagonal gauge invariant object (\ref{GI2}), built with the field strength tensor off-diagonal tetrad components, is also zero locally. It is consistent because this object is precisely invariant under $SU(2)$ local gauge transformations (except for a global negative sign in one particular discrete case, which makes no difference if the object is zero). Then, we conclude, if its components are all null, zero in one gauge, they all will be null in any other gauge. The two remaining blocks associated to the two remaining locally gauge invariant objects in the diagonal of the field strength tensor, cannot be diagonalized by any tetrad rotations in the planes one and two through the use of the gauge vectors $X_{d}^{\sigma}$ and $Y_{d}^{\sigma}$. That is, by $SU(2)$ tetrad gauge transformations on these planes, that have been proven to be equivalent to tetrad Lorentz transformations LB1 and LB2 on these planes. This happens because under LB1 or LB2 local gauge transformations, the local gauge invariant objects (\ref{GI1}) and (\ref{GI3}) are not only invariant, but they are composed with only one tetrad component each, which makes them invariant as single objects and therefore, if locally they are not zero in one gauge they are not zero in any other gauge. It is evident that the ``block diagonal gauge'' might be a source of simplification in dealing with the field equations, and of course the inherent simplification in the geometrical analysis of any problem involving these kind of fields (\ref{eyme}-\ref{ymvfe2}).

\section{Application: Observables found with the new tetrads}
\label{application:observables}

In order to find observables in the non-Abelian case we are going to follow the pattern established by the Abelian case. From equation (\ref{EMF}) we notice the existence of two local observables ($\sqrt{-Q/2}\:\:\cos\alpha, \sqrt{-Q/2}\:\:\sin\alpha$). These two local scalars are on one hand coordinate invariant, and on the other, gauge invariant under the local group $U(1)$ of electromagnetic gauge transformations \cite{A}. The Abelian situation is automatic because the diagonalization is automatic. The non-Abelian case however requires of the techniques introduced in the previous sections \ref{Gauge invariants} and \ref{diagonal}. Therefore, let us introduce three orthogonal unit local vectors in isospace $\vec{s} = s^{a}\:\sigma^{a}$, $\vec{t} = t^{a}\:\sigma^{a}$ and $\vec{u} = u^{a}\:\sigma^{a}$ described in terms of local cosines as in expressions (\ref{ISO}-\ref{ISOSUM}). Let us block diagonalize the field strength independently in all three isospace directions. For the sake of notational simplicity we write,

\begin{eqnarray}
Tr[\vec{s}\: \cdot \: f_{\mu\nu}] &=& -2\:\sqrt{-Q_{s}/2}\:\:A_{s}\:\:\:\overline{S}_{(o)[\mu}\:\overline{S}_{(1)\nu]} +
2\:\sqrt{-Q_{s}/2}\:\:B_{s}\:\:\overline{S}_{(2)[\mu}\:\overline{S}_{(3)\nu]} \label{fprojS} \\
Tr[\vec{t}\: \cdot \: f_{\mu\nu}] &=& -2\:\sqrt{-Q_{t}/2}\:\:A_{t}\:\:\:\overline{T}_{(o)[\mu}\:\overline{T}_{(1)\nu]} +
2\:\sqrt{-Q_{t}/2}\:\:B_{t}\:\:\overline{T}_{(2)[\mu}\:\overline{T}_{(3)\nu]} \label{fprojT} \\
Tr[\vec{u}\: \cdot \: f_{\mu\nu}] &=& -2\:\sqrt{-Q_{u}/2}\:\:A_{u}\:\:\:\overline{U}_{(o)[\mu}\:\overline{U}_{(1)\nu]} +
2\:\sqrt{-Q_{u}/2}\:\:B_{u}\:\:\overline{U}_{(2)[\mu}\:\overline{U}_{(3)\nu]} \ .\label{fprojU}
\end{eqnarray}

It is clear then, that it is possible to express the field strength in general as,

\begin{eqnarray}
\vec{f}_{\mu\nu} &=& \vec{s}\:\:Tr[\vec{s}\: \cdot \: f_{\mu\nu}] + \vec{t}\:\:Tr[\vec{t}\: \cdot \: f_{\mu\nu}] + \vec{u}\:\:Tr[\vec{u}\: \cdot \: f_{\mu\nu}]
\ .\label{fgenexpression}
\end{eqnarray}

This expression is general, since the three unit orthogonal local isovectors $\vec{s}$, $\vec{t}$ and $\vec{u}$ are arbitrary, it just suffices for them not to be trivial and be described through equations like (\ref{ISO}-\ref{ISOSUM}). The important point here is that the three new local objects given by $Tr[\vec{s}\: \cdot \: f_{\mu\nu}]$, $Tr[\vec{t}\: \cdot \: f_{\mu\nu}]$ and $Tr[\vec{u}\: \cdot \: f_{\mu\nu}]$ are invariant under local spatial three dimensional rotations in isospace. Which is equivalent to say, under local non-Abelian $SU(2)$ gauge transformations, see sections \ref{Gauge invariants} and \ref{diagonal}. In the local unit iso-surface. Therefore, we have found six local observables. They are the three pairs ($\sqrt{-Q_{s}/2}\:\:A_{s}, \sqrt{-Q_{s}/2}\:\:B_{s}$), ($\sqrt{-Q_{t}/2}\:\:A_{t}, \sqrt{-Q_{t}/2}\:\:B_{t}$) and ($\sqrt{-Q_{u}/2}\:\:A_{u}, \sqrt{-Q_{u}/2}\:\:B_{u}$). These three pairs of local objects have the following properties. They are locally invariant under general coordinate transformations. They are locally invariant under $SU(2) \times U(1)$ gauge transformations, see reference \cite{A2}. Therefore, they are local observables. It remains to be studied the Dirac-Bergmann nature of these new local observables, task that is not trivial according to references \cite{AK}$^{-}$\cite{LLIV}.

\section{Conclusions}
\label{conclusions}

We have been able to develop a new local gauge invariant method under the group $SU(2)$ for the block diagonalization of a $SU(2)$ field strength tensor. Projected in any possible direction in the local isospace. It is evident that any problem involving the field strength tensor will be maximally simplified. It is also evident that we can extend this method to the local $SU(3)$ case. In this latter case $SU(2)$ would just be a local subgroup. In the process we found three new local gauge invariant objects built with the field strength and our new tetrads. These new local gauge invariant tetrad objects helped us understand why we can block diagonalize in a local gauge invariant way. These new tetrads reveal the link between local gauge symmetries and gravitational symmetries of Yang-Mills theories in four-dimensional Lorentzian spacetimes and help find the true geometrical degrees of freedom. Simplification through the use of symmetries, that these new tetrads help understand with clarity. It is also clear that along with the local gauge invariant method developed in manuscript \cite{A2} in order to diagonalize the stress-energy tensor, we have with this new local gauge invariant method to maximally simplify the field strength, a new gauge invariant method to classify Yang-Mills field theories \cite{WY}. As an application, we found new local observables. It remains to be studied as a future application, their Dirac-Bergmann nature, see references \cite{AK}$^{-}$\cite{LLIV}. We quote from \cite{EW} ``The relationship between theoretical physics and geometry is in many ways very different today than it was just ten or fifteen years ago. It used to be that when one thought of geometry in physics, one thought chiefly of classical physics-and in particular of general relativity-rather than quantum physics. Geometrical ideas seemed (except perhaps to some visionaries) to be far removed from quantum physics-that is, from the bulk of contemporary physics. Of course, quantum physics had from the beginning a marked influence in many areas of mathematics-functional analysis and representation theory, just to mention two. But it would probably be fair to say that twenty years ago the day to day preoccupations of most practicing theoretical elementary particle physicists were far removed from considerations of geometry''.

\section{Appendix I}
\label{sec:appI}

This appendix is introducing the object $\Sigma^{\alpha\beta}$. This object according to the matrix definitions introduced in the references is Hermitic. The use of this object in the construction of our tetrads allows for the local $SU(2)$ gauge transformations $S$, to get in turn transformed into purely geometrical transformations. That is, local rotations of the $U(1)$ electromagnetic tetrads $E_{\alpha}^{\:\:\rho}$ included in the definitions of $X^{\sigma} = Y^{\sigma} = Tr[\Sigma^{\alpha\beta}\:E_{\alpha}^{\:\:\rho}\: E_{\beta}^{\:\:\lambda}\:\ast \xi_{\rho}^{\:\:\sigma}\:\ast \xi_{\lambda\tau}\:A^{\tau}]$.
The object $\sigma^{\alpha\beta}$ is defined as $\sigma^{\alpha\beta} = \sigma_{+}^{\alpha}\:\sigma_{-}^{\beta}-\sigma_{+}^{\beta}\:\sigma_{-}^{\alpha}$, \cite{MK}$^{,}$\cite{GM}. The object $\sigma_{\pm}^{\alpha}$ arises when building the Weyl representation for left handed and right handed spinors. According to \cite{GM}, it is defined as $\sigma_{\pm}^{\alpha} = (\bf{1},\pm\sigma^{i})$, where $\sigma^{i}$ are the Pauli matrices for $i = 1\cdots3$. Under the $(\frac{1}{2},0)$ and $(0,\frac{1}{2})$ spinor representations of the Lorentz group it transforms as,

\begin{equation}
S_{(1/2)}^{-1}\:\sigma_{\pm}^{\alpha}\:S_{(1/2)} = \Lambda^{\alpha}_{\:\:\:\gamma}\:\sigma_{\pm}^{\gamma}\ .\label{sigmatr}
\end{equation}

Equation (\ref{sigmatr}) means that under the spinor representation of the Lorentz group, $\sigma_{\pm}^{\alpha}$ transform as vectors. In (\ref{sigmatr}), the matrices $S_{(1/2)}$ are local, as well as $\Lambda^{\alpha}_{\:\:\:\gamma}$ \cite{GM}. The $SU(2)$ elements can be considered to belong to the Weyl spinor representation of the Lorentz group. Since the group $SU(2)$ has a homomorphic relationship to $SO(3)$, they just represent local space rotations. It is also possible to define the object $\sigma^{\dagger\alpha\beta} = \sigma_{-}^{\alpha}\:\sigma_{+}^{\beta}-\sigma_{-}^{\beta}\:\sigma_{+}^{\alpha}$, analogously. Then, we have,

\begin{center}
$\imath \: \left(\sigma^{\alpha\beta} + \sigma^{\dagger\alpha\beta}  \right)  = \left\{ \begin{array}{ll}
				0 \:\:\:\:\: \mbox{if $\alpha = 0$ and $\beta = i$}\\
				4\:\epsilon^{ijk}\:\sigma^{k} \:\:\:\:\: \mbox{if $\alpha = i$ and $\beta = j$ \ ,}
				    \end{array}
			    \right. $
\end{center}

\begin{center}
$ \sigma^{\alpha\beta} - \sigma^{\dagger\alpha\beta}  = \left\{ \begin{array}{ll}
				-4\:\sigma^{i} \:\:\:\:\: \mbox{if $\alpha = 0$ and $\beta = i$}\\
				0 \:\:\:\:\: \mbox{if $\alpha = i$ and $\beta = j$ \ .}
				    \end{array}
			    \right. $
\end{center}

We might then call $\Sigma_{ROT}^{\alpha\beta} = \imath \: \left(\sigma^{\alpha\beta} + \sigma^{\dagger\alpha\beta} \right)$, and $\Sigma_{BOOST}^{\alpha\beta} = \imath \: \left(\sigma^{\alpha\beta} - \sigma^{\dagger\alpha\beta} \right)$. Therefore, a possible choice for the object $\Sigma^{\alpha\beta}$ could be for instance $\Sigma^{\alpha\beta} = \Sigma_{ROT}^{\alpha\beta} + \Sigma_{BOOST}^{\alpha\beta}$. This is a particularly suitable choice when we consider proper Lorentz transformations of the tetrad vectors nested within the structure of the gauge vectors $X^{\mu}$ and $Y^{\mu}$. For spatial, that is, rotations of the $U(1)$ electromagnetic tetrad vectors which in turn are nested within the structure of the two gauge vectors $X^{\mu}$ and $Y^{\mu}$, as is the case under study in this paper, we can simply consider $\Sigma^{\alpha\beta} = \Sigma_{ROT}^{\alpha\beta}$. These possible choices also ensure the Hermiticity of gauge vectors.  Since in the definition of the gauge vectors $X^{\mu}$ and $Y^{\mu}$ we are taking the trace, then $X^{\mu}$ and $Y^{\mu}$ are real.


\end{document}